\documentclass[sigconf,nonacm,screen]{acmart}
\AtBeginDocument{%
  \providecommand\BibTeX{{%
    \normalfont B\kern-0.5em{\scshape i\kern-0.25em b}\kern-0.8em\TeX}}}

\acmConference[EAAMO '23]{Equity and Access in Algorithms, Mechanisms, and Optimization }{October 30--November 01,
  2023}{Boston, MA}
\acmPrice{15.00}
\acmISBN{978-1-4503-XXXX-X/18/06}

\usepackage{bm}
\usepackage{mathtools}

\newcommand{\recsum}[1]{\ensuremath{R^{#1}}}

\begin{document}

\title[Deploying a Robust Active Preference Elicitation Algorithm on MTurk]{Deploying a Robust Active Preference Elicitation Algorithm on MTurk: Experiment Design, Interface, and Evaluation for COVID-19 Patient Prioritization}

\author{Caroline M. Johnston}
\email{cmjohnst@usc.edu}
\affiliation{%
  \institution{USC Center for AI in Society}
  \city{Los Angeles}
  \state{California}
  \country{USA}
}

\author{Patrick Vossler}
\email{pvossler@usc.edu}
\affiliation{%
  \institution{USC Center for AI in Society}
  \city{Los Angeles}
  \state{California}
  \country{USA}}

\author{Simon Blessenohl}
\email{blesseno@usc.edu}
\affiliation{%
  \institution{USC School of Philosophy}
  \city{Los Angeles}
  \state{California}
  \country{USA}}

\author{Phebe Vayanos}
 \email{phebe.vayanos@usc.edu}
\affiliation{%
  \institution{USC Center for AI in Society}
  \city{Los Angeles}
  \state{California}
  \country{USA}
}


\begin{abstract} 
Preference elicitation leverages AI or optimization to learn stakeholder preferences in settings ranging from marketing to public policy. The online  robust preference elicitation procedure of \cite{Vayanos2020} has been shown in simulation to outperform various other elicitation procedures in terms of effectively learning individuals' true utilities.
However, as with any simulation, the method makes a series of assumptions that cannot easily be verified to hold true beyond simulation. 
Thus, we propose to validate the robust method's performance using real users, 
focusing on the particular challenge of selecting policies for prioritizing COVID-19 patients for scarce hospital resources during the pandemic. To this end, we develop an online platform for preference elicitation where users report their preferences between alternatives over a moderate number of pairwise comparisons chosen by a particular elicitation procedure.
We recruit 193 Amazon Mechanical Turk (MTurk) workers to report their preferences and demonstrate that the robust method outperforms asking random queries by $21\%$, the next best performing method in the simulated results of~\cite{Vayanos2020}, in terms of recommending policies with a higher utility.
\end{abstract}






\maketitle

\section{Introduction}\label{sec:Introduction} 
The difficulty in eliciting the true preferences of individuals  is well documented~\cite{Green1978}. This can occur due to various factors such as the need to elicit over a large feature space or inconsistencies when individuals report their preferences. More generally, it is difficult for individuals to express their utility exactly or in terms of specific magnitude, though it is generally agreed that, inconsistencies aside, individuals can reliably report preferences in the form of ordinal rankings. 

One type of preference elicitation is conjoint analysis, where AI or optimization is used to design surveys to learn stakeholders' preferences or value judgements. Though many conjoint analysis methods attempt to address these aforementioned issues, few have been validated beyond simulation in a setting with real users. Specifically, the online robust method of \cite{Vayanos2020} demonstrates strong performance in simulation in its ability to elicit individuals' preferences in settings with large feature spaces (up to 20 features) and inconsistent responses compared to other methods in the literature. 

Many works involving participatory design and preference elicitation arise within social good settings, where there is an inherent tension or trade-off between fairness and efficiency, see e.g.,~\cite{Bertsimas2012}. For example, a group of policymakers may desire to design policies that match kidney donors to patients awaiting a transplant~\cite{Freedman2020, McElfresh2021}, policies that allocate scarce resources to individuals experiencing homelessness~\cite{Vayanos2020}, or models for recidivism prediction~\cite{Yu2020}. Specifically in~\cite{Vayanos2020}, policymakers may need to decide between implementing a policy that assigns resources such that the number of individuals that exit homelessness is maximized overall, a policy that equalizes this outcome across different racial groups, and a policy(ies) that makes some compromise between these two outcomes. In each of these settings, an effective preference elicitation method determines which policy or model is most preferred by a policymaker as they consider these difficult trade-offs. Once this is determined, the method ultimately facilitates the decision that can be implemented in reality.
 
In this work we develop an online interface for preference elicitation that evaluates the effectiveness of asking the online robust queries of~\cite{Vayanos2020} compared to asking randomly selected queries, the next best performing elicitation procedure in the authors' simulated experiments. 
Using our interface, we recruit workers from Amazon Mechanical Turk\footnotemark \footnotetext{\url{https://www.mturk.com/}} (MTurk) to report their preferences between COVID-19 patient prioritization policies. Our contributions are as follows:
1) We design a preference elicitation platform implementing the approach in~\cite{Vayanos2020} to elicit preferences over COVID-19 patient prioritization policies using real data from the United Kingdom and recruit participants to use the platform through MTurk. 2) We conduct rigorous analysis to demonstrate that the robust method is much more likely to recommend a preferred policy to users in this real setting 
compared to  asking random queries. Thus, we validate in an experimental setting with real users 
that the robust method is more effective at  determining individuals' true preferences.

The rest of the paper is structured as follows. In Section \ref{sec:LitReview}, we review the current state of the literature. In Section \ref{sec:PreferenceElicitationModel}, we summarize the robust preference elicitation and recommendation optimization model of \cite{Vayanos2020}. We then discuss our COVID-19 patient prioritization setting in Section \ref{sec:COVID19PatientPrioritization} and our experimental design and results in Sections \ref{sec:ExperimentalDesign} and \ref{sec:Results}, respectively. Finally, we  discuss the limitations of the model and experiment in  Section \ref{sec:Limitations} and the ethical impacts of designing and implementing such a system in Section \ref{sec:EthicalImpact}.

\section{Literature Review}\label{sec:LitReview}
 \paragraph{Preference Elicitation} 
 Preference elicitation has been used in a plethora of AI applications from combinatorial auctions \cite{Sandholm2006, Brero2021}, to homeless services \cite{Kuo2023}, to food donation transportation \cite{Lee2019}, to content moderation \cite{Shen2022}. Within healthcare, researchers have explored preference elicitation for kidney exchange \cite{Freedman2020} and  services provided at rheumatology outpatient clinics \cite{Ryan2001}. Many works investigate different ways of incorporating preferences into AI systems. For example, in \cite{Freedman2020}, participants are asked an exhaustive set of queries to elicit their preferences which are then encoded into a kidney exchange matching algorithm while \cite{Ryan2001} uses discrete choice experiments. Additionally, many preference elicitation works take a human-computer interaction lens, often using qualitative methods. Both \cite{Lee2019} and \cite{Kuo2023} use interviews with real study participants to assess their opinions concerning specific algorithms in their respective domains. \cite{Shen2022} conducts workshops in which groups collaboratively discuss which content moderation algorithm should be implemented within their community.
 
\paragraph{Conjoint Analysis}
 Conjoint analysis is a specific type of preference elicitation. In many conjoint analysis works, the decision-maker is able to query the user (agent) for their preferences for particular policies (items or alternatives) within a limited time frame or with a limited budget. Thus, the goal is to learn the user's true preferences (utility) by optimizing the selection of queries presented to the user. With these preferences, a suitable policy recommendation can then be made to the user. 
 
Many works assume that the utilities of agents take a convex~\cite{Armbruster2015} or additive form~\cite{Toubia2003,Toubia2004,Boutilier2006,OHair2013,Vayanos2020, Braziunas10}, facilitating the use of convex optimization tools for preference elicitation. One stream of the literature takes a set-based approach to the uncertainty in an agent's true utility while optimizing over various types of queries.
 For example,~\cite{Toubia2004, Bertsimas2012} optimize the selection of pairwise comparisons of the form ``Do you prefer item A over item B?'' while~\cite{Toubia2003} optimizes queries that ask the agent to report the strength of their preferences between two items, i.e. ``By \textit{how much} do you prefer item A over item B?'' Taking a slightly different approach,~\cite{ Boutilier2006} optimizes the selection of gamble queries of the form ``Do you prefer item A or a gamble in which you receive item A with probability $p$ and item~B with probability~$1-p$?'' Though varying in the types of queries, each of these works select their respective queries such that they reduce the largest amount of uncertainty in the agent's true utility. Focusing on minimizing the regret of the recommended item, \cite{Braziunas10} uses a heuristic to select both pairwise comparison and gamble queries. 

 Another stream of the literature takes a Bayesian approach to the uncertainty in the agent's true utility~\cite{Chajewska2000, Boutilier2002, Zhao2018, Pfeiffer2012, Saure2019}. In this setting, the uncertainty is represented by a prior distribution which is updated as the agent responds to queries. For example,~\cite{Chajewska2000} and~\cite{Pfeiffer2012} select gamble and pairwise comparison queries, respectively, such that they maximize the expected information gain with respect to the agent's true utility.  
 
We now discuss the online robust preference elicitation method of~\cite{Vayanos2020}, which selects queries that maximize the worst-case utility of the recommended item using set-based uncertainty. This method is appealing in terms of both its design and performance. For example, many of the works mentioned above focus solely on selecting queries in order to learn the true preferences of individuals. However, in many settings, the ultimate goal is to make  a suitable recommendation to the individual, not necessarily an exact determination of their true utility. The authors address this by integrating the selection of queries and the recommendation as a single robust optimization problem. 
The method additionally accounts for inconsistencies in individuals' responses, modeling them as idiosyncratic shocks in utility that are normally distributed. 
In terms of the performance of the method, robust optimization is often criticized for being over-conservative. However, the authors demonstrate through simulation that the method performs well relative to the individuals' \textit{true} utility, in addition to its realization in the worst-case, outperforming various methods in the literature. However, as noted in Section \ref{sec:Introduction}, it is often difficult for individuals to specify their true utility exactly, whereas in simulation these true utilities can be generated and accessed at will.  
Thus, it remains to be seen
whether these positive results hold in a setting with real users. 
\paragraph{Evaluation of Preference Elicitation Algorithms with Real Users} 
Of the conjoint analysis works above,~\cite{Toubia2003,Toubia2004,Toubia2007,Braziunas10} validate their findings in settings with real users. In \cite{Toubia2003}, the authors compare their polyhedral conjoint analysis method to other online elicitation procedures for recommending laptop-computer bags 
~($n=330$). They validate their method's accuracy in  estimating true utilities through randomly chosen holdout queries and the user's selection of a most preferred bag from 5 randomly selected bags.  To validate their probabilistic polyhedral method, \cite{Toubia2007} focuses on wine consumers $(n=2255)$. They use a similar experimental design as above except each participant answers queries selected by \textit{both} the proposed method and a competing method. \cite{Toubia2004} tests their polyhedral method with real users to design an executive education program ($n=354$), additionally examining how quickly their method converges to its final utility estimation compared to a competing method. The validation of the heuristic method of \cite{Braziunas10} 
    involves asking users for an exhaustive listing of their preferences over 100 apartments ($n=40$). Users then report their preferences over the union of the highly rated apartments according to their method and the highly rated apartments according to the exhaustive preference listing. 
    
    Our own experimental design to evaluate the robust method of \cite{Vayanos2020} is most similar to the validation in \cite{Toubia2007}  where we require users to answer both queries chosen by the robust method and queries chosen randomly. This design, as the authors of \cite{Toubia2007} note, is favorable due to the ability to compare methods within users. Thus, we can more concretely determine which method demonstrates better performance independent of potential idiosyncrasies of a given individual.

  Perhaps the most wide-scale AI preference elicitation experiment is the ``Moral Machine,'' where real users report their preferences in ethical dilemmas faced by autonomous vehicles~\cite{Awad2018MoralMachine}. The platform presents participants 
  with pairwise comparisons of scenarios with randomly selected features who must choose how the autonomous vehicle behaves. Methodologically, the authors take a different approach to those above. They do not assume any functional form of the utilities of agents and take a causal inference perspective, attempting to identify the causal effects of the randomly selected features of the scenarios.
The authors find that, at least at the time of the study, individuals' preferences were in direct conflict with the governmental policy guidelines proposed by the German Ethics Commission on Automated and Connected Driving. Thus, this work 
provides a real-life example and strong motivation for 
 eliciting preferences in a general public policy setting. By understanding how government guidelines may conflict with the population's value judgements for such issues, policymakers can begin to understand how to better mitigate such conflicts.

 \section{Preference Elicitation Model}\label{sec:PreferenceElicitationModel}
 
In this paper, we 
validate the method  of \cite{Vayanos2020}, which for the rest of this work we will refer to as \texttt{Robust}, on real users from MTurk. We present a high-level description of the model here in an effort to keep this work self-contained. We refer the reader to the work itself for more detail.  

Given a finite set of alternatives, the goal of \texttt{Robust} is to recommend an alternative to an agent whose true utility is unknown. The method can gain (noisy) information about the agent's true utility by eliciting their preferences through a moderate number of queries. These queries are pairwise comparisons of the form ``Do you prefer alternative A or alternative B?'' The method uses concepts from robust optimization to optimize the selection of the alternatives within the queries to recommend an alternative that maximizes the agent's worst-case utility.

\subsection{Utility and Query Model}\label{sec:PrefUtilityModel} 
Let $\mathcal{X} \subseteq \mathbb{R}^J$ be the set of alternatives from which \texttt{Robust} asks queries and selects a recommendation, indexed in the set~$\mathcal{I}=\{1,\dots,I\}$. 
In \texttt{Robust}, each ${\bm x} \in \mathcal{X}$ is modeled as a vector of~$J$ real-valued attributes. In our COVID-19 resource allocation setting, we are interested in eliciting preferences for policies that prioritize patients for treatment. 
We represent these policies as a vector of $J$ efficiency and fairness metrics. In other words, this vector encodes the values of policy outcomes that policymakers may take into account when making 
such decisions (see Section~\ref{sec:COVIDPolicyGeneration}). 
 
 In \texttt{Robust}, a query is a comparison between two alternatives. 
 The set of possible queries is $\mathcal{C} := \left \{ (i, i') \in \mathcal{I}^2 \mid i < i' \right \}$. A particular choice of $K$ queries indexed in the set $\mathcal{K} := \{1,\dots,K \}$ is represented by~$\bm \iota \in {\mathcal{C}}^K$, which specifies which alternatives are compared in the~$K$ queries. For $\kappa \in \mathcal{K}$,  $\bm \iota^{\kappa} := (\bm \iota_{1}^\kappa, \bm \iota_{2}^\kappa) \in \mathcal{C}$ denotes that the~$\kappa$th query elicits the preference between alternatives $\bm x^{\bm \iota_{1}^\kappa}$ and~$\bm x^{\bm \iota_{2}^\kappa}$. 

\texttt{Robust}  assumes that each agent's utility function is linear in $\bm x\in\mathbb{R}^J$. 
 Therefore, for an alternative $\bm x \in \mathcal{X}$, the method represents the agent's utility for that alternative as $\bm u^\top \bm x$. In this setting, the true utility of the agent is uncertain and~$\mathcal{U}\subseteq \mathbb{R}^{J}$ is used to denote the agent's initial uncertainty set, 
 where each element ${\bm u} \in{\mathcal{U}}$ represents one possible realization of the agent's utility function. The method assumes that $\mathcal{U}$ is a non-empty bounded polyhedron such that~$\mathcal{U} = \left \{ \bm u \in \mathbb{R}^{J} \mid \bm B \bm u \geq \bm b \right \}$, for some $\bm B \in \mathbb{R}^{M \times J}$,~$\bm b \in \mathbb{R}^{M}.$
 
When asked the $\kappa$th query, an agent is able to respond in one of three ways using the elements of $\mathcal{S} := \left \{ -1, 0, 1 \right \}$: either the agent prefers alternative 1 ($\bm s_{\kappa} =1$), is indifferent between the alternatives~($\bm s_{\kappa} =0$), or prefers alternative 2 ($\bm s_{\kappa} = -1$). \texttt{Robust} assumes that these responses are inconsistent or noisy and
defines the parameter $\Gamma$ as a threshold of inconsistency in the agent's responses. These inconsistencies are assumed to lie in the set~$\mathcal{E}_{\Gamma} := \{\bm \epsilon \in \mathbb{R}^{K}_{+} : \sum_{\kappa \in \mathcal{K}} \bm \epsilon_{\kappa} \leq \Gamma\}$ and that the elements of $\bm \epsilon \in \mathbb{R}_{+}^{K}$ are independent and normally distributed with standard deviation $\sigma$. 

For each $\kappa \in \mathcal{K}$, once the agent responds to query $\bm \iota^{\kappa}$ with response~$\bm s_{\kappa}$, \texttt{Robust} updates  the uncertainty set for the agent's utility according to
\[
\mathcal{U}_{\Gamma(\kappa)}( \overline{\bm \iota}^{\kappa}, \overline{\bm s}_{\kappa}) := \left \{
\begin{array}{ccc}
\bm u \in \mathcal{U}: \exists \,\, \bm \epsilon \in \mathcal{E}_{\Gamma(k)} \text{  s.t. } \forall \, k \in \{1, \dots, \kappa\} \\ 
{\bm u}^\top \left (\bm x^{\bm \iota_{1}^{k }} - \bm x^{\bm \iota_{2}^{k}} \right ) \geq - \bm \epsilon_{k} : \bm s_k  = 1 \\  
\left \vert \, {\bm u}^\top \left (\bm x^{\bm \iota_{1}^{k }} - \bm x^{\bm \iota_{2}^{k }} \right ) \,\right \vert  \leq \bm \epsilon_{k } :\bm s_k = 0 \\
{\bm u}^\top \left (\bm x^{\bm \iota_{1}^{k }} - \bm x^{\bm \iota_{2}^{k}} \right) \leq \bm \epsilon_{k }    : \bm s_k  = -1 
\end{array} \right \},
\]
where $\overline{\bm \iota}^{\kappa} := \{\bm \iota^k\}_{k=1}^{\kappa}$ and $\overline{\bm s}_{\kappa}:= \{\bm s_k\}_{k=1}^{\kappa}$, denote the $\kappa$ queries and responses, respectively, that have been elicited thus far and  $\Gamma(k)$ denotes the level of inconsistency at query $k\in \{1, \dots, \kappa\}$. By the assumptions on $\bm \epsilon$, we have that~$\Gamma(k) = 2\sigma \sqrt{k}\,\text{erf}^{-1}(2p-1)$ for $k \in \{1 ,\dots, \kappa \}$, ensuring that $\bm \epsilon \in \mathcal{E}_{\Gamma(k)}$ with probability $p$.

\subsection{Preference Elicitation Optimization}\label{sec:PreferenceElicitationOptimizationModel}

For a set of alternatives $\mathcal{X}$ and agent's initial uncertainty set $\mathcal{U}$, \texttt{Robust} aims to solve the following \textit{robust recommendation problem} to select an alternative that will maximize the agent's worst-case utility
\begin{equation}
\tag{\recsum{\mathcal{U}}}
z :=  \max_{{\bm x} \in \mathcal{X}}  \; \min_{{\bm u} \in \mathcal{U}} \;  {\bm u }^\top{\bm x}.
  \label{eqn:recc-problem}
\end{equation}

As mentioned above, the method is able to query the agent and update its knowledge of the agent's true utility in the uncertainty set relative to these queries and respective responses. This is referred to as \textit{online} elicitation, in which queries are selected one at a time. As each query is selected and the agent provides their response, the method integrates this information into the uncertainty set and uses it to adaptively select the next query. This online elicitation procedure will ask more informative queries and lead to a recommendation with higher worst-case utility than those chosen offline, or chosen all at once before receiving any of the agent's responses, as shown in \cite{Vayanos2020}.

\texttt{Robust} solves
 a series of optimization problems for each $\kappa$ for $\kappa = 1, \dots, K$, updating the uncertainty set as it elicits the agent's preference at each query. 
Specifically, for each $\kappa$, the method solves the following problem
\begin{equation}
\max_{\bm \iota^{\kappa} \in \mathcal{C}} \; \min_{{\bm s_\kappa} \in \mathcal{S}} \; \max_{{\bm x} \in \mathcal{X}}  \; \min_{{\bm u} \in \mathcal{U}_{\Gamma(\kappa)}(\overline{\bm \iota}^{\kappa}, \overline{{\bm s}}_{\kappa})} \; {\bm u}^\top {\bm x},
\label{eq:online-approx}
\end{equation} 
which can be reformulated as a finite mixed-binary linear program (see \cite{Vayanos2020} for details). Note that in the problem above, $\{\bm \iota^k\}^{\kappa-1}_{k=1}$ and~$\{\bm s_k\}^{\kappa-1}_{k=1}$ are data according to the queries and responses that have been previously elicited.
As formulated above, the method  is robust with respect to both the worst-case response to the selected query  ${\bm s}_{\kappa} \in \mathcal{S}$ and the realization of the agent's utility coefficients~${\bm u} \in \mathcal{U}_{\Gamma(\kappa)}(\overline{\bm \iota}^\kappa, \overline{\bm s}^\kappa)$ as characterized by the selected queries and respective responses. Finally, we point out that Problem \eqref{eq:online-approx}  is a deterministic process. Thus, we can compute each optimal query~$\bm \iota^{\kappa} \in \mathcal{C}$ for every possible sequence of previous responses~$\{\bm s_k\}^{\kappa-1}_{k=1}$ that a user can give. 
We will leverage this in our experimental design (see Section \ref{sec:Procedue}).

\section{COVID-19 Patient Prioritization }\label{sec:COVID19PatientPrioritization} 
We will evaluate \texttt{Robust} through the problem of designing policies that allocate scarce critical care unit (CCU) beds during the COVID-19 pandemic, using real data from the United Kingdom. Throughout the pandemic, many hospitals have become inundated with an influx of patients, leading to resource shortages of life-saving equipment such as ventilators and CCU beds~\cite{Ranney2020}. Doctors must then decide which patients receive this scarce equipment and who must go without. Not only does this impose an emotional burden on doctors and medical staff~\cite{Greenberg2020,Ferraresi2020}, but without a disciplined way of allocating these resources, this can likely lead to an inefficient, unfair, or inconsistent allocation. Therefore, hospitals implement policies that assign these scarce resources to patients in times of overwhelming need. There currently exist many candidate policies for prioritizing patients for resources, from first-come-first-served, to point scoring rules based on the severity of disease, to a prediction of probability of hospital death (see e.g.,~\cite{Ezekiel2020,Rapsang2014} for surveys and discussion). Each of these policies has unique performance outcomes and ethical implications that must be considered when implementing them in a real hospital. Thus, the use of a preference elicitation tool on stakeholders such as doctors can bring great value in determining the allocation policy that these stakeholders most prefer.

\section{Experimental Design}\label{sec:ExperimentalDesign}

We now describe the experiment we have designed to evaluate the effectiveness of \texttt{Robust} in a  
 setting with real users from MTurk.
We recruit participants to report their preferences in a questionnaire of pairwise comparisons chosen by \texttt{Robust} and randomly selected pairwise comparisons.
As the final query in the questionnaire, users directly report their preference between the robustly recommended alternatives according to each procedure, determining which is more effective at recommending an item that the user prefers. 

\subsection{Participants and Incentives}\label{sec:Participants}
 We recruited~250 participants from MTurk to take our questionnaire. We collected participants' demographic information such as age group, race/ethnicity, gender, and whether the individual works in healthcare or not. All participants were required to be at least 18 years old, have access to the Internet, and be proficient in English. Participants were only allowed to take the survey once with a time limit of two hours. They were paid $\$2.5$ once they finished answering all pairwise comparisons in the questionnaire. Our study is IRB approved through an exempt review.

 \subsection{Procedure} \label{sec:Procedue}
 \paragraph{Pairwise Comparisons}
 To successfully complete the questionnaire, each participant answers $2K +1$  pairwise comparison queries. 
 Each user is randomly assigned to one of two groups. The first group answers $K$  queries that are chosen by
 \texttt{Robust} as described in Section~\ref{sec:PreferenceElicitationOptimizationModel}, a set of ``memory-cleansing'' questions (see below), and then $K$ queries that are chosen randomly. The second group takes the survey in the opposite order, first answering the randomly selected queries, ``memory-cleansing'' questions, and then the queries chosen by \texttt{Robust}. For each query, we randomize the order of its presentation, i.e., which alternative will appear on the left side of the screen versus the right side of the screen. This randomization of placement on the screen and the random assignment of users to each group offsets the presence of order biases that has been well-studied in choice-based conjoint analysis, see e.g. \cite{Chrzan1994}. 
 
Regardless of which group the user is placed in, once they have completed the $2K$  queries described above, we then compute the recommended alternatives for the user according to Problem \recsum{\mathcal{U}} relative to the selected queries and responses for each method. 
More concretely, let $\bm \iota^{\texttt{robust}} \in  \mathcal{C}^K$ and $\bm s^{\texttt{robust}}\in\mathcal{S}^K$ be the queries chosen by \texttt{Robust} and corresponding responses, respectively. We solve Problem~$\recsum{\mathcal{U}_{\Gamma}(\bm \iota^{\texttt{robust}}, \bm s^{\texttt{robust}})}$ to obtain $\bm x^*_{\texttt{robust}} \in \mathcal{X}$, the optimal alternative to recommend to the user with respect to the preference information elicited by \texttt{Robust}, and $ z^*_{\texttt{robust}} \in \mathbb{R}$, the worst-case utility guaranteed by this recommendation. We similarly solve Problem~$\recsum{\mathcal{U}_{\Gamma}(\bm \iota^{\texttt{rand}}, \bm s^{\texttt{rand}})}$ to obtain $\bm x^*_{\texttt{rand}}\in \mathcal{X}$ and~$ z^*_{\texttt{rand}} \in \mathbb{R}$ for the queries selected randomly and corresponding responses, respectively.  

The $(2K+1)$th, or final, pairwise comparison we present to the user is a comparison between $\bm x^*_{\texttt {robust}}$ and~$\bm x^*_{\texttt {rand}}$
where we similarly randomize the presentation of the alternatives. Since each alternative is the optimal recommendation under each elicitation procedure, this final comparison enables us to directly determine which of the two methods recommends the item that is preferred.

\paragraph{Memory-cleansing and Attention-checking}
Memory-cleansing questions play an important role in our experiment design by allowing us to directly compare the performance of both methods on each user.
 For this process, users answer the following three questions, known as the ``Cognitive Reflection Test''~\cite{Frederick2005}:
 \begin{enumerate}
\item A bat and a ball cost \$1.10 in total. The bat costs \$1.00 more than the ball. How much does the ball cost?
\item If it takes 5 machines 5 minutes to make 5 widgets, how long would it take 100 machines to make 100 widgets?
\item In a lake, there is a patch of lily pads. Every day, the patch doubles in size. If it takes 48 days for the patch to cover the entire lake, how long would it take for the patch to cover half of the lake?
 \end{enumerate}
 These questions serve two purposes. First, by having the users focus on this unrelated task before switching from one part of the questionnaire to the other, these questions help to reduce the presence of a carry-over effect from method to method. 
 Secondly, we require the users to answer these questions as free-form text as a form of attention-checking. Note that we do not check for correctness of the users' answers but rather that the answers are ``sensible'', e.g., not a random string of characters or other ``bot-like'' behavior. We also record the time it took for the users to answer each pairwise comparison query for \textit{post hoc} attention-checking analysis (see Section \ref{sec:PreferencePreProcess}).

\paragraph{\texttt{Robust} Query and Recommendation Calculation}
We recall that, the formulation of Problem \eqref{eq:online-approx} is a deterministic process and that we can compute each optimal query for every possible sequence of responses that a user can give before running the experiment. By storing these optimal queries in a lookup table, 
 this further reduces the potential bias between the two query selection methods. Now each query that is presented to the user can be shown ``instantaneously'' whether it is selected by \texttt{Robust} or randomly. In other words, there is no need to wait for \texttt{Robust} to solve its optimization problem to select the next query to ask. 
 
To create the lookup table we solve the mixed-binary linear programming formulation of Problem \eqref{eq:online-approx}  
for $K=10$ and with initial partworth uncertainty set  $
 \displaystyle \mathcal{U} =  \left \{ \bm u \in \mathbb{R}_{+}^{J} : {\bm u}^\top\textbf{e} = 1 \right \}$, as in the experimental results of~\cite{Vayanos2020}. We solve this for every possible sequence of responses $\{\bm s_k\}^{\kappa-1}_{k=1}$
 that a user can report for $\kappa \in \mathcal{K}$. We generate each $\Gamma(k)$ according to the formula given in Section~\ref{sec:PrefUtilityModel} with $p=90\%$ and $\sigma=0.1$, representing a 90\% confidence level that the $\bm \epsilon_{k}$ lie within $\mathcal{E}_{\Gamma(k)}$. We use the same $p$ value as~\cite{Vayanos2020} but choose a slightly larger $\sigma$ value than what the authors present. We use this larger $\sigma$ value since we cannot validate whether a choice of $\sigma$ is misspecified without access to the user's true utilities, the difficulty of which in an 
experimental setting with real users is discussed in Sections \ref{sec:Introduction} and~\ref{sec:LitReview}. We note that by using a higher value of $\sigma$, we decrease the chance of misspecifying the model of \texttt{Robust} at the risk of obtaining over-conservative results. Nevertheless, we demonstrate that, even if this choice of $\sigma$ is over-conservative, \texttt{Robust} performs well in this setting (see Section \ref{sec:AnalysisOfResults}). To conclude, with the parameters above, we solve Problem \eqref{eq:online-approx} for each  $3^1 + 3^2 +\dots+3^9$ 
 possible sequence of responses. For each user's respective queries and responses, we solve the robust recommendation problems
 $\recsum{\mathcal{U}_{\Gamma}(\bm \iota^{\texttt{robust}}, \bm s^{\texttt{robust}})}$ and $\recsum{\mathcal{U}_{\Gamma}(\bm \iota^{\texttt{rand}}, \bm s^{\texttt{rand}})}$ with an initial partworth uncertainty set and~$\Gamma=\Gamma(10)$. All problems are solved on 6 cores of Xeon 2.6GHz cores and 1GB of memory using Gurobi~10.0.0.

\section{Results}\label{sec:Results}
Using our platform, study participants take a questionnaire where they report their preferences between prioritization policies for COVID-19 patients. These policies are generated using real data from the United Kingdom as in~\cite{Johnston2020}.
 For each user, we evaluate if the robust recommendation under the preference information obtained by \texttt{Robust} is preferred compared to the robust recommendation under the preference information obtained by randomly selected queries.

 \subsection{COVID-19 Patient Prioritization Policy Generation}\label{sec:COVIDPolicyGeneration}
  We now describe how we use real data to simulate policies for prioritizing COVID-19 patients for treatment which we use as the alternatives in our experiment. We note that one could evaluate a preference elicitation method on randomly generated vectors of policy features. However, evaluating on feature vectors that are the outcomes of implementable policies computed using real data will arguably yield stronger evidence about the method's performance in real-world settings. Additionally,  this differentiates our work from discrete choice experiments used in healthcare \cite{Ryan2001}. In a discrete choice experiment, features and their levels are combined and individuals are asked their preferences between such combinations in pairwise comparisons. However, it is entirely possible that no policy exists that achieves such combinations of  features, especially if these features are policy outcomes. Thus, by simulating policies and using these simulated outcomes as features, we can obtain an actionable policy that aligns with individuals' preferences.

  We create a simulation to estimate the outcomes of 25 counterfactual COVID-19 patient prioritization policies by age group at the country-level using data from the United Kingdom from April 1st to July 15th,~2020.  As estimates of 
  the rate of daily arrivals of patients in need of critical care, we use the United Kingdom's daily projections of the expected number of COVID-19 CCU patients over this time period from the Institute for Health Metrics and Evaluation (IHME) model.\footnotemark \footnotetext{\url{http://www.healthdata.org/covid/data-downloads}} As estimates of 
  both the distribution of these patients' ages and their probability of recovery, we use the historical proportions and survival rates of COVID-19 patients in the CCU by age as provided by the UK Intensive Care National Audit and Research Centre\footnotemark \footnotetext{\url{https://www.icnarc.org/Our-Audit/Audits/Cmp/Reports}} over this time period.
 Note that we use only age as a patient characteristic  due  to the unavailability of patient outcome data based on, say, both age and race.
  
  A policy assigns a score to each patient based on their age and the number of days they have waited for a critical care unit (CCU) bed, allocating beds to the highest-scoring patients. We consider policies in the form of regression trees. For a given policy, we calculate a patient's score by starting at the root node and traversing the tree, proceeding to the left or right child of a node depending on whether the patient satisfies the given condition until a leaf node is reached. A leaf node contains a real number between 0 and 1, representing the patient's score. We generate 25 regression tree policies of depth three by randomly picking a feature and a comparison value for each branching node and a random number between 0 and 1 for the leaf nodes. These leaf node values create a priority ranking for individuals for treatment 
  based on patient characteristics, for which this general concept is used in many CCUs (see Section \ref{sec:COVID19PatientPrioritization}). 

For a given prioritization policy, we simulate events in the following order for each day in the time frame of our simulation:~1)\ First, COVID-19 patients who require CCU beds arrive at a hospital(s) according to our estimated daily arrival rate;~2)\ Some patients in critical care pass away or recover according to our estimated survival rate that is dependent on their age, making these CCU beds newly available;~3)\ Some waiting patients, who did not receive a bed, pass away according to a constant probability independent of their age;~4)\ Finally, the policy assigns waiting patients to available (finite) CCU beds, making these CCU beds unavailable to other patients. If at this point there are no patients waiting for a bed, no patients in the CCU, and no expected future arrivals of patients, we terminate the simulation.

The features we estimate, and by which users will evaluate policies, are:  1) the total number of life-years saved, 2) the overall survival probability of patients, 3) the survival probabilities of patients across six age groups, and 4) the probabilities of patients receiving a CCU bed across these same six age groups. We also  record the coefficients of variation (CV) with respect to metrics 3) and 4) in order to capture the notion that individuals may prefer more ``fair'' policies, i.e., policies with low CV values. We choose these metrics to represent various efficiency and fairness metrics that policymakers may consider in such a setting, where 1) and 2) are efficiency metrics and 3) and 4) are fairness metrics with respect to age. Specifically, 3) represents fairness in terms of outcomes and 4) represents fairness in terms of access. Thus, the trade-offs between these metrics must be considered when users report their preferences. Domain experts may include other metrics of interest as appropriate.

Because the CV value is less interpretable compared to the other features, we do not show this metric directly to the user within the pairwise comparisons. However, since \texttt{Robust} and the robust recommendations will have access to these values, the methods may still determine how this notion of fairness contributes to the user's utility. In total, each of the~25 randomly generated policies is characterized by 16 features, a similar scale in number of alternatives and features as tested in~\cite{Vayanos2020}.
These~25 policies become our set of alternatives $\mathcal{X}$, for which there are~$300$ unique pairwise comparisons that each user could be asked.

  \begin{table*}
  \caption{Demographics of Study Participants ($n=193$)}
  \label{tab:Turker-demographics}
  \begin{tabular}{lc| lc| lc}
    \toprule
    Demographic&Frequency (\%) &Demographic&Frequency (\%) &Demographic&Frequency (\%)  \\
    \midrule
    White & 82 & 18-39 years of age & 78  &  High school or equivalent& 18  \\
    Asian & 11 & 40-49 years of age & 16 &     Some college or more & 82 \\
    Black/African American & 3 & 50-69 years of age & 6 &  Works in healthcare & 49 \\
    Hispanic/Latino & 3&  Male & 61  &Does not work in healthcare & 51  \\
    American Indian/Alaska Native & 1 & Female & 39 & & \\ 
  \bottomrule
\end{tabular}
\end{table*}

\subsection{Preference Elicitation Platform}\label{sec:PrefElicitationPlatform}
\begin{figure}[t]
  \centering
\includegraphics[width=\linewidth]{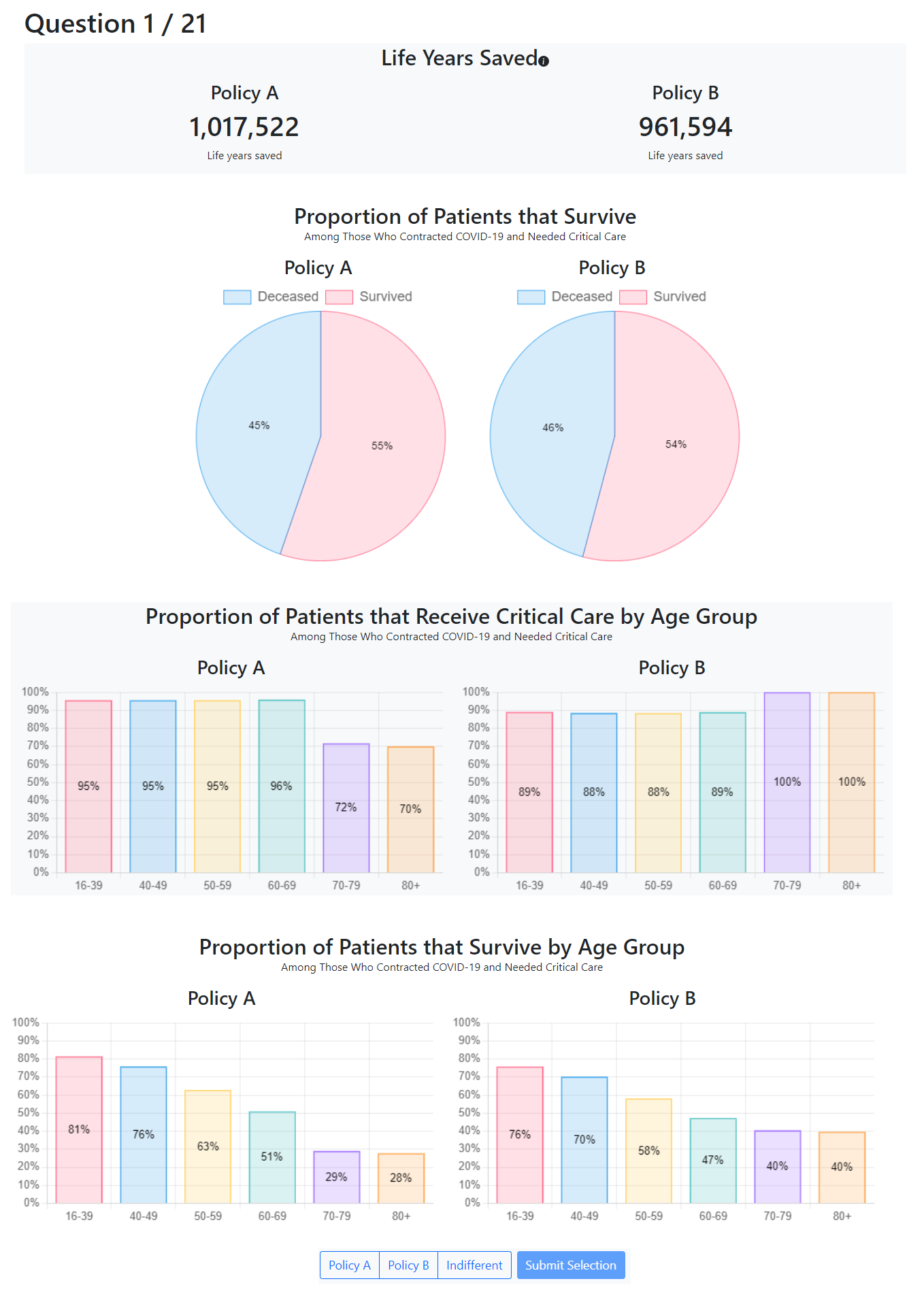}
  \caption{An example of a pairwise comparison between two COVID-19 patient prioritization policies displayed in the interface.} \Description[Interface pairwise comparison example]{The interface web-page with a two-column design. The page displays the features of two COVID-19 patient prioritization policies, one policy in each column. The features are "Life Years Saved", presented as raw values; "Proportion of Patients that Survive", presented as a pie chart; "Proportion of Patients that Receive Critical Care by Age Group", presented as a bar chart; and "Proportion of Patients that Survive by Age Group", presented as a bar chart. At the bottom of the page, there are buttons to select and submit the preferences of the user.}
  \label{fig:example-query}
\end{figure}

We designed a platform\footnotemark \footnotetext{\url{https://www.cais-preference-elicitation.com/}} to  
evaluate \texttt{Robust} on MTurk over a period of two consecutive days in an effort to diversify the sample of the study population.
After participants provide consent, they view the landing page of the questionnaire. This page familiarizes them with our setting, asking them to imagine that they are a healthcare professional working at a hospital during May of 2020 -- before the wide-scale availability of vaccines and resources for treating COVID-19. We instruct them that their goal is to help determine a set of guidelines at a hospital to decide which patients will receive a bed, ventilator, or other lifesaving treatment in a CCU, when there are more patients than available resources. They are also given a high-level overview of the outcomes upon which they will evaluate the policies. An example of a pairwise comparison in the platform and the outcomes of policy features is seen in Figure \ref{fig:example-query}.

\subsection{User Preference Data Pre-Processing}\label{sec:PreferencePreProcess}
We now describe our data cleaning procedure. For any survey responses where the user took the survey multiple times, though they were instructed not to, we kept their first attempt but removed later attempts to avoid potential bias in these repeated exposures. We also manually checked and removed any users who provided  bot-like answers to the memory-cleansing questions (see Section~\ref{sec:Procedue}). We removed any users that took less than 15 seconds to answer the first query or that took less than an average of 3 seconds to answer the following queries as \textit{post hoc} attention checks.
We characterize these using two different cut-off values because we observed that users take a longer amount of time on the first query to become familiar with the structure of our questionnaire and interface design. As the survey continues, they may answer queries more quickly as they become more familiar with the process. 
 We also removed an outlier user that took more than one hour to complete the survey as no other user took more than 40 minutes to complete the survey. 
 We note that our analysis below remains significant when filtering out participants who took less than 30 seconds to answer the first query and less than 5 seconds on average  to answer the remaining queries. Introducing larger cut-off values removed too many samples to be able to make any statistical claims such as those that follow. 

  Finally, we note that it is possible that the final query in the questionnaire between $\bm x^*_{\texttt {robust}} \in \mathcal{X} $ and $\bm x^*_{\texttt {rand}} \in \mathcal{X}$ is such that the two policies are the same, i.e., $\bm x^*_{\texttt {robust}} = \bm x^*_{\texttt {rand}}$. From \texttt{Robust}'s construction of queries in $\mathcal{C}$ and the manner in which we select random queries, this is the only time in which a query with the same two policies can be presented to a user. When this occurs, this means that the preference information gained from asking robust queries and asking random queries is equivalent insofar as the same policy will maximize the user's worst-case utility. We remove any responses in which this occurs and the user does not report that they are indifferent between the two policies as a form of attention-checking, since these two policies are exactly the same.

 After we clean the data as described above, our final population size is 193 participants. We can see the demographic information of this group in Table \ref{tab:Turker-demographics}. Our population skews toward young, white males with at least some college education. These types of biases are well-documented for MTurk workers, see e.g., \cite{Berinsky2012, Ross2010}.  The participants took an average of 6.5 minutes with a standard deviation of 5.5 minutes to complete the questionnaire. 
 
\subsection{Analysis of User Preference Data} \label{sec:AnalysisOfResults}
We analyze the unique preferences of the users, which is noteworthy for two reasons. First,  many elicitation methods evaluated in simulation assume that utilities are randomly distributed over the population~\citep{OHair2013, Saure2019, Vayanos2020}. Because our study population lacks demographic diversity (see Table \ref{tab:Turker-demographics}), it is important to determine if there is also a lack of diversity in preferences. Second, if the questionnaire reveals that a majority of users' preferences are the same, developing an online preference elicitation tool is moot. We could find a recommended policy that is ``one-size-fits-all,'' no matter the user. 

\begin{figure}[t]
  \centering
\includegraphics[width=0.75\linewidth]{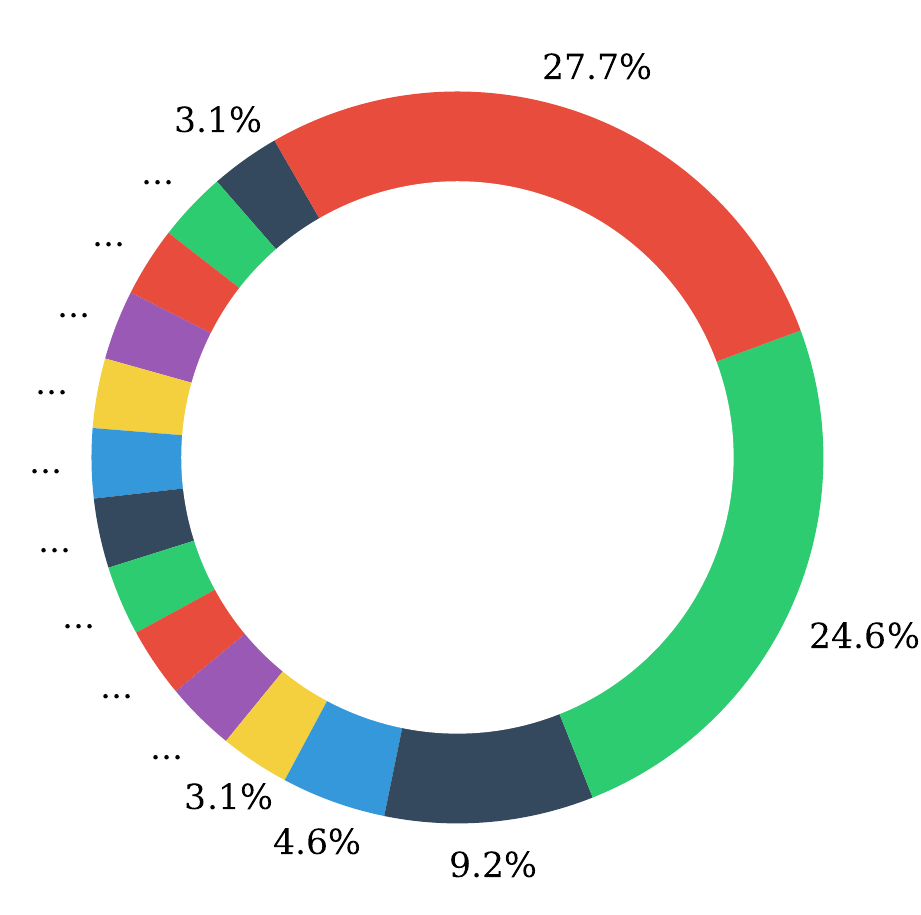}
  \caption{The percentage of users who had the same preferences as at least one other user in the study ($n=65$) for the queries chosen by \texttt{Robust}. Each sector refers to a unique set of preferences, $\bm s^{\texttt{robust}}\in\mathcal{S}^K $. 
  } \Description[Pie chart of users who have same preference as at least one other user]{A pie chart displaying the percentage of users who share the same preferences as at least one other user in the study for the robust queries. There are 15 sectors of the graph, where the largest sectors are 27.7 percent and 24.6 percent of the population (n=65). The majority (11) of the sectors are 3.1 percent of the population.}
  \label{fig:survey-unique-preferences2}
\end{figure}

To determine the diversity of preferences, we recall that the robust portion of the questionnaire is the same for those that report the same preferences. Thus, a subset of users can be asked the same sequence of queries~$\bm \iota^{\texttt{robust}} \in  \mathcal{C}^K$ by \texttt{Robust} if they report the same sequence of responses~$\bm s^{\texttt{robust}}\in\mathcal{S}^K$. Of the 193 users, 128 had unique preferences and, therefore, answered a unique set of queries selected by \texttt{Robust}. In Figure~\ref{fig:survey-unique-preferences2}, we show a breakdown of the remaining~65 users who shared the same preferences with at least one other user for the queries selected by \texttt{Robust}. Though some subsets of the study population have the same preferences, a majority differ in this regard.

\begin{figure}[t]
  \centering
\includegraphics[width=\linewidth]{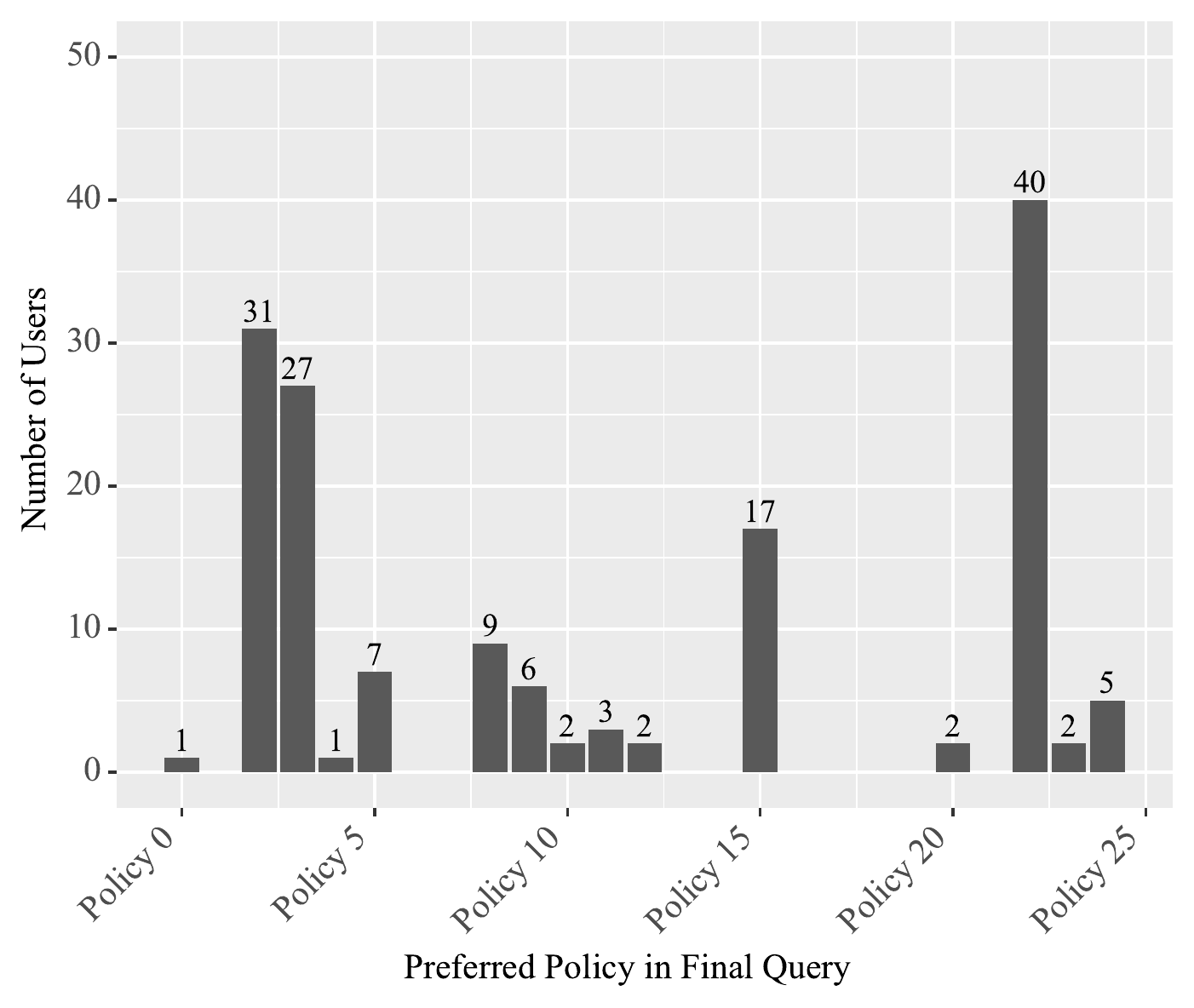}
  \caption{
  The number of users who select the given policy as their  preferred policy in the last query of the questionnaire, which directly compares the policy recommended by \texttt{Robust} and the policy recommend by asking random queries~($n=155$). Note that we exclude those that are indifferent in the last query.  } \Description[Bar chart of users' preferred policy in final query]{A bar chart showing number of users from 0 to 50 on the Y axis against the preferred policy in the final query for policy 0 to policy 24 on the X axis. The most preferred policies are policy 2,3, and 22 which are preferred by 31, 27, and 40 users, respectively. 12 other policies are preferred by smaller numbers of users.} 
  \label{fig:survey-policy-preferences}
\end{figure}

Even though users may have differing preferences with respect to the individual queries selected by \texttt{Robust}, this may be less relevant if the policy that we recommend to these users is the same. In Figure \ref{fig:survey-policy-preferences}, we report the number of users who prefer each policy in the last query of the questionnaire. Note that we exclude those that  are indifferent in this final query. Though there are a small number of policies that many users prefer in the final query, overall, there is a high amount of variability in these responses. Thus, we can reasonably conclude that because of the diversity of preferences of the users, designing and  
using such an elicitation and recommendation procedure is of practical importance.

\begin{figure}[t]
  \centering
\includegraphics[width=\linewidth]{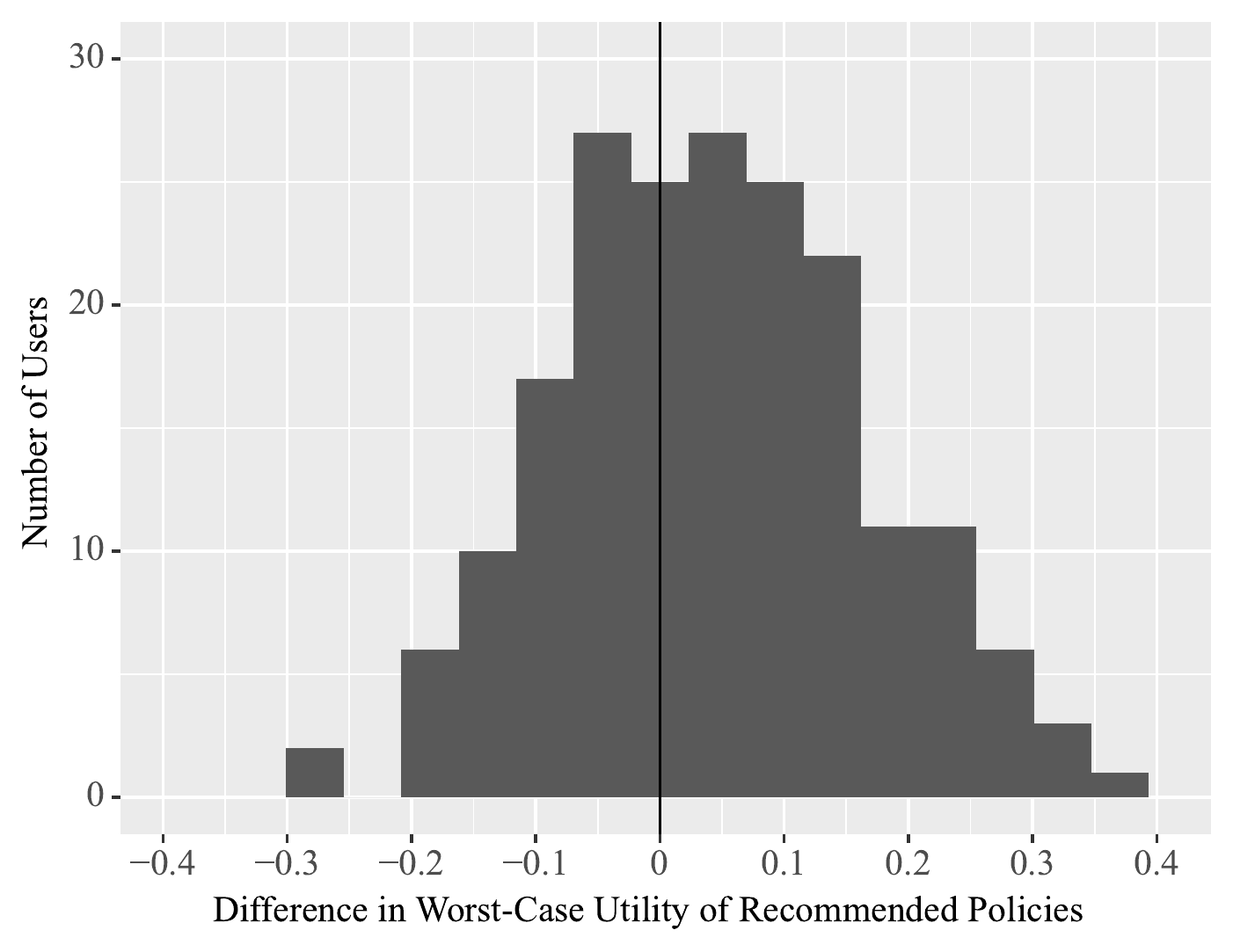}
  \caption{The distribution of the difference in the worst-case partworth utilities of the policy recommended by \texttt{Robust} and the policy recommended by asking random queries ($n=193$). Any user with a positive difference value will have 
 a higher utility in the worst-case for~$\bm x^*_{\text{robust}}$ than for $ \bm x^*_{\text{rand}}$. The difference in utility values is normalized between $-1$ and $1$, where a value of $1 (-1)$ corresponds to $\bm x^*_{\text{robust}} (\bm x^*_{\text{rand}})$ as the recommended policy if the utility is known, i.e., the best possible recommendation, and $\bm x^*_{\text{rand}} (\bm x^*_{\text{robust}})$ as the recommended policy when asking no queries, i.e., the worst possible recommendation.}  \Description[Histogram of user differences in worst-case utilities for recommended policies]{A histogram of number of users from 0 to 30 on the Y axis and the difference in worst-case utilities for the robust method and random method on the X axis from -0.4 to 0.4 in increments of 0.1. The majority of responses are clustered in the range of -0.1 to 0.15. There are more users that have a value greater than 0 than less than 0.}
  \label{fig:diff-in-wc-util}
\end{figure}

We recall that \texttt{Robust} selects queries that optimize the worst-case utility of the recommended item. We additionally emphasize that both $\bm x^*_{\texttt{robust}}$ and $\bm x^*_{\texttt{rand}}$ are selected to maximize the worst-case utility of the user with respect to the uncertainty set as characterized by the queries and responses obtained by each method. Thus, even though random queries are selected, the recommendation of~$\bm x^*_{\texttt{rand}}$ is made in a robust manner. In Figure \ref{fig:diff-in-wc-util}, we investigate the distribution of the difference in the worst-case utilities between the recommended policies $\bm x^*_{\texttt{robust}}$ and $\bm x^*_{\texttt{rand}}$ for each user as characterized by their respective uncertainty sets, i.e., the value of $z^*_{\texttt{robust}} -  z^*_{\texttt{rand}}$.  \texttt{Robust} recommends a policy with an average worst-case utility of~0.60, while asking random queries recommends a policy with an average worst-case utility of 0.52. This difference in average guarantee of utility is statistically significant, $t(384)= 4.58, p < 0.001$ and demonstrates the benefit of \texttt{Robust} when optimizing for this worst-case performance.

\begin{figure}[t]
  \centering
\includegraphics[width=\linewidth]{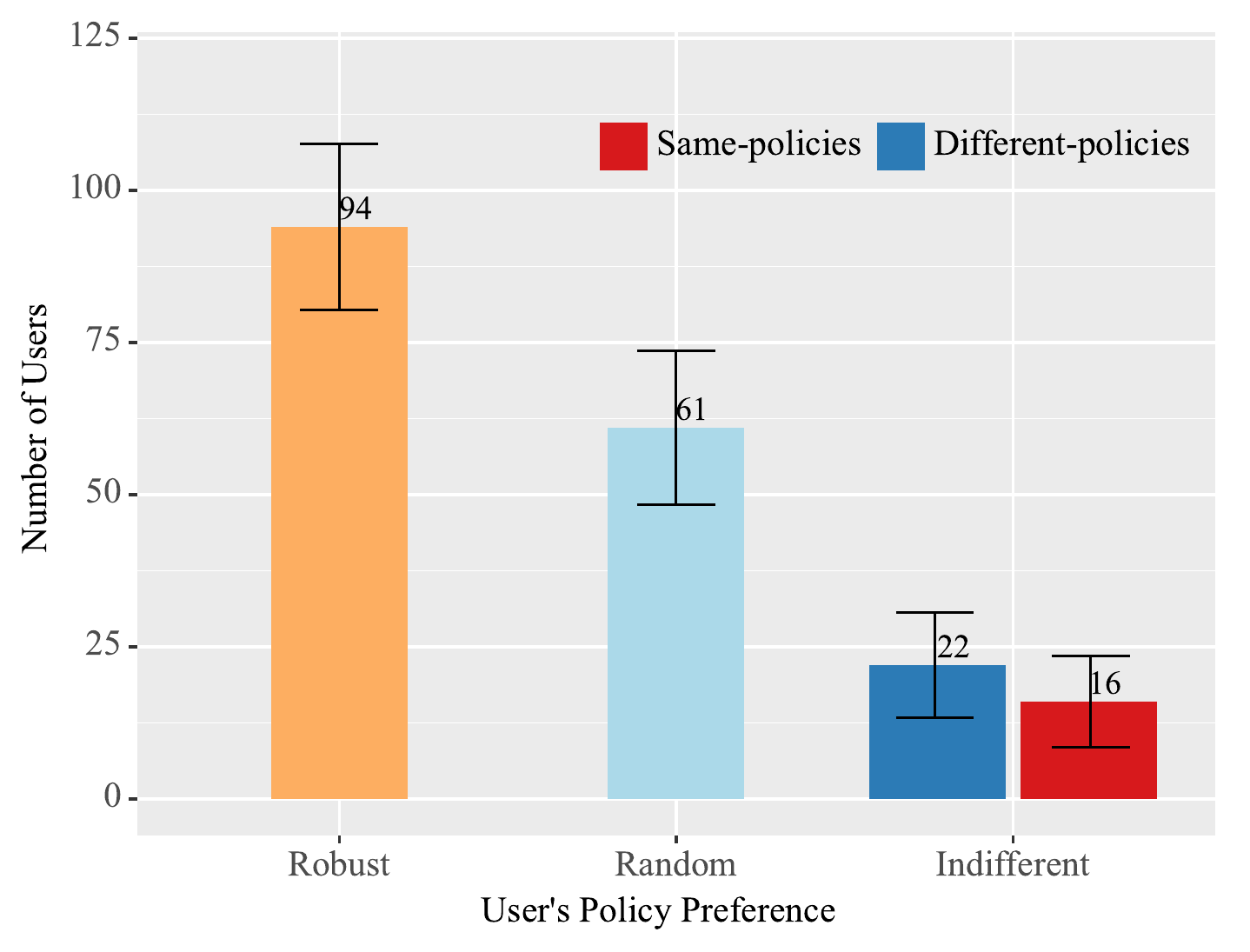}
  \caption{
  Results of the last query of the questionnaire which directly compares the policy recommended by \texttt{Robust} and the policy recommend by asking random queries ($n=193$) with $95\%$ confidence intervals. \texttt{Robust} (\texttt{Random}) is the number of users that selected that they prefer $\bm x^*_{\texttt {robust}}(\bm x^*_{\texttt {rand}})$. \texttt{Indifferent} refers to the number of users that reported that they are indifferent. \texttt{Different-policies} (\texttt{Same-policies}) refers to the instances where a user is presented with policies~$\bm x^*_{\texttt {robust}}\neq \bm x^*_{\texttt {rand}}$ ( $\bm x^*_{\texttt {robust}}=\bm x^*_{\texttt{rand}}$) and reports an indifference between the two.  
  }\Description[Bar chart of user responses in final query]{A bar chart showing the number of users from 0 to 125 on the Y axis with 95 percent confidence intervals (CI) against those who reported that they prefer the policy recommended by the robust method, recommended by asking random queries, or are indifferent between the two on the X axis. The indifferent responses are additionally reported between those who are indifferent when shown the same two policies in the final query and those that are shown two different policies. There are 94 users that preferred the robust policy (CI from about 80 to about 108), 61 users who preferred the random policy (CI from about 48 to 74), 22 who are indifferent when shown different policies (CI from about 13 to 31), and 16 who are indifferent when shown the same policies (CI from about 8 to about 24).}
  \label{fig:survey-results}
\end{figure}

Ultimately, we desire favorable performance in terms of the user's true utility, or how they actually report their preference. Figure \ref{fig:survey-results} displays the number of users who report that they prefer $\bm x^*_{\texttt {robust}}$, the policy that \texttt{Robust} recommends; prefer~$\bm x^*_{\texttt {rand}}$, the policy that asking random queries recommends; or are indifferent between the two policies. As mentioned in Section \ref{sec:PreferencePreProcess}, it is possible that $\bm x^*_{\texttt {robust}}$ is the same as  $\bm x^*_{\texttt {rand}}$. Thus, we disaggregate users who report that they are indifferent by whether these policies are the same or not. By disaggregating, we can differentiate between when the two methods make an equivalent recommendation in terms of the policies themselves versus an equivalent recommendation in terms of the user's utility for two unique policies. 
 
 We first note that indifferences are reported between the two policies for ${\sim}20\%$ of the users, i.e., the two methods are equivalent in their recommendations for ${\sim}20\%$ of users by either recommending the same policy or two policies with similar utility. When considering users that are not indifferent in the final query ($n=155)$, \texttt{Robust} is able to recommend a  policy that is more preferred  for $33$, or ${\sim}21\%$, more users compared to asking queries randomly.

We validate that we cannot detect various order effects in our results. We cannot detect a statistically significant difference in the results shown in Figure \ref{fig:survey-results} between the users who first answered the queries selected by \texttt{Robust} $(n=94)$ versus those that first answered the queries selected randomly $(n=99)$, $t(191) = 0.80, p = 0.43$. 
We similarly do not have enough statistical evidence to believe there is a bias in our results concerning whether, in the final query,~$\bm x^*_{\texttt {robust}}$ appears on the right side $(n=76)$ or the left side $(n=117)$ of the screen,~$t(191) = 1.47, p = 0.14$.

Finally, we validate that our results in Figure \ref{fig:survey-results} are not due to randomness. Specifically, we compare our results to that of users uniformly selecting their preference between the policies displayed in the final query. We focus on our results that have strict preferences, i.e., that are not indifferent ($n=155$). This is justifiable since we observe that users are unlikely to report that they are indifferent between any two policies. Out of the 21 total queries asked to users across all questionnaires, users reported an indifference for only $9\%$ of the queries. In fact, only 46\% of users reported any indifference at all during the questionnaire, while the rest always reported strict preferences. Therefore, we do not believe there exists uniformity between the three choices and only investigate uniformity when selecting a strict preference between the two policies. 
To this end, we compare our results to that of a uniform distribution, or users selecting between the two policies in the final query at random, and find that our result is statistically significant and not due to randomness,~$\chi^2(1,155) = 6.61,p < 0.01$. Thus, we have significant evidence that \texttt{Robust} will lead to robust recommendations that users are more likely to prefer, or that have higher utility, compared to the robust recommendations that result from asking random queries.

\section{Limitations}\label{sec:Limitations}
The ideal audience for our interface and application as described in Section \ref{sec:COVID19PatientPrioritization} is stakeholders in the healthcare field, not necessarily members of the general public. Though about half of the participants reported that they work in healthcare (see Table \ref{tab:Turker-demographics}), the ones that do not may have lacked the necessary domain knowledge to report their preferences in a faithful or meaningful way. However, we tried to familiarize all users with our particular setting through the information provided in the platform's landing page (see Section~\ref{sec:PrefElicitationPlatform}).

Our study evaluates whether \texttt{Robust} can  recommend an alternative with higher utility compared to the recommended alternative when asking random queries. We make no claim as to whether, for example, either alternative is the user's top-ranked alternative. This would require an exhaustive listing of the user's preferences, a time-intensive process, or for the user to be able to characterize their utility exactly, which is highly impractical as discussed in Sections \ref{sec:Introduction} and \ref{sec:LitReview}.

\section{Ethical Impacts}\label{sec:EthicalImpact}
Through our study, we have shown that using the online robust method of \cite{Vayanos2020} results in a recommendation to an individual that they are more likely to prefer compared to asking queries randomly. However, as with the adoption of any AI system, the end-users of such a method must be accepting of any potential shortcomings in the method's performance. From our results in Figure \ref{fig:survey-results}, there certainly exist users for which \texttt{Robust} does not recommend a preferred policy. However, if this method is still an improvement over stakeholders' status quo procedure for policy design and implementation, then it may be reasonable to adopt it. Furthermore, comparison of  \texttt{Robust} to other preference elicitation methods in the literature in an 
experimental setting with real users is a sensible area for future exploration.  

Any recommendation system faces the potential issue of automation bias, where individuals are more likely to trust in decisions proposed by AI models as opposed to humans, see e.g., \cite{Goddard2012}. To mitigate this risk in a real decision-making setting, we suggest that the policy this system recommends serves as a starting point for further refinement, as appropriate, and not necessarily strict adoption.

Finally, the elicitation and recommendation system is limited to the features that are explicitly provided in the alternatives. In our COVID-19 resource allocation policy setting in Section \ref{sec:COVIDPolicyGeneration}, there may exist other latent policy features concerning efficiency, fairness, or equity that are not stated as features of the alternatives but have real consequences if the recommended policy is implemented in reality. One way to address this is to have an initial elicitation process in which stakeholders report their preferences for what features should be represented in the alternatives \cite{Freedman2020}. Once these features are determined, the appropriate alternatives can be generated to find an agent's recommendation.

\section{Conclusion}
 
In this work, we develop an interface for validating the online robust preference elicitation method of \cite{Vayanos2020} in settings with real users.  Focusing on an application of designing COVID-19 policies that assign scarce resources to patients, we investigate this method compared to asking random queries using workers from MTurk. We find that these robust queries recommend a preferred policy for~$21\%$ more users compared to the policy that is recommended when making a robust recommendation with randomly selected queries. Thus, we validate that the robust method is an effective tool for eliciting individuals' preferences in an 
experimental setting with real users beyond that of simulation.

\begin{acks}
    C.M. Johnston acknowledges support under the NSF GRFP (NSF Grant Number DGE-1842487). P. Vayanos and P. Vossler acknowledge support under the USC Zumberge Special Solicitation – Epidemic \& Virus Related Research and Development award.  This material is based upon work supported by the National Science Foundation under CAREER Grant No. 2046230. P.\ Vayanos gratefully acknowledges this support. The authors thank the authors of~\cite{Vayanos2020} for their code and the members of the USC Center for AI in Society who gave initial feedback on the interface. The authors acknowledge the Center for Advanced Research Computing (CARC) at the University of Southern California for providing computing resources that have contributed to the research results reported within this publication. URL: \url{https://carc.usc.edu}. 
\end{acks}


\bibliographystyle{ACM-Reference-Format}
\bibliography{references}


\end{document}